\documentclass[10pt]{article}
\usepackage{latexsym,graphicx}
\newcommand{\be}{\begin{equation}}
\newcommand{\ee}{\end{equation}}
\def\n{\noindent}
\catcode `\@=11
 \catcode `\@=12
\begin{document}
\begin{center}
\large{\bf{An Interacting Two-Fluid Scenario for Dark Energy in FRW Universe}} \\
\vspace{10mm}
\normalsize{Hassan Amirhashchi$^1$, Anirudh Pradhan$^2$ and Bijan Saha$^3$}\\
\vspace{5mm}
\normalsize{$^{1}$Department of Physics, Islamic Azad University, Mahshahr Branch, Mahshahr, Iran \\
E-mail: hashchi@yahoo.com; h.amirhashchi@mahshahriau.ac.ir} \\
\vspace{5mm}
\normalsize{$^{2}$Department of Mathematics, Hindu Post-graduate College, Zamania-232 331,
Ghazipur, India \\
E-mail: pradhan@iucaa.ernet.in; acpradhan@yahoo.com} \\
\vspace{5mm}
\normalsize{$^{2,3}$Laboratory of Information Technologies, Joint Institute for Nuclear Research,
141980 Dubna, Russia \\
$^3$E-mail: bijan@jinr.ru } \\
\end{center}
\vspace{10mm}
\begin{abstract}
We study the evolution of the dark energy parameter within the scope of a spatially flat and isotropic 
Friedmann-Robertson-Walker (FRW) model filled with barotropic fluid and dark energy. To obtain the deterministic 
solution we choose the scale factor $a(t) = \sqrt{t e^{t}}$ which yields a time dependent deceleration parameter 
(DP). In doing so we consider the case minimally coupled with dark energy to the perfect fluid as well as direct 
interaction with it.
\end{abstract}
 \smallskip
\n Keywords : FRW universe, Dark energy, Variable deceleration parameter \\
\n PACS number: 98.80.Es, 98.80-k, 95.36.+x
\section{Introduction}
Observations of distant Supernovae (SNe Ia) \cite{ref1}$-$\cite{ref6}, fluctuation of cosmic microwave background 
radiation (CMBR) \cite{ref7,ref8}, large scale structure (LSS) \cite{ref9,ref10}, sloan digital sky survey (SDSS) 
\cite{ref11,ref12}, Wilkinson microwave anisotropy probe (WMAP) \cite{ref13} and Chandra x-ray observatory 
\cite{ref14} by means of ground and altitudinal experiments have shown that our Universe is spatially flat 
and expanding with acceleration. This fact can be put in agreement with the theory if one assumes that the 
Universe is basically filled with so-called dark energy. The measurement of photometric distances to the 
cosmological Supernova, supported by a number of independent arguments, in particular by the observational 
data on the angular temporal fluctuations of CMBR, shows that the lion share of the energy density of matter 
belongs to non-baryonic matter. This form of matter cannot be detected in laboratory and does not interact 
with electromagnetic radiation. Given the fact that almost three fourth of energy density of the Universe 
originated from dark energy and plays crucial role in the accelerated mode of expansion of the Universe,
there appear a large number of models capable of describing this dark energy.\\

Dark energy models with higher derivative terms were constructed by Zhang and Liu \cite{ref15}. The cosmological
evolution of a two-field dilaton model of dark energy was investigated by Liang et al. \cite{ref16}. Viscous dark
energy models with variable $G$ and $\Lambda$ were studied by Arbab \cite{ref17}. The modified Chaplygin gas with
interaction between holographic dark energy and dark matter was investigated in Ref. \cite{ref18}. The tachyon 
cosmology in interacting and non-interacting cases in non-flat FRW Universe was studied in Ref. \cite{ref19}. In 
this Letter we study the evolution of the dark energy parameter within the framework of a FRW cosmological model 
filled with two fluids. In doing so we consider both interacting and non-interacting cases. 
\section{The Metric and Field  Equations}
We consider the spatial homogeneous and isotropic Friedmann-Robertson-Walker (FRW) metric as
\begin{equation}
\label{eq1}
ds^{2} = -dt^{2} + a^{2}(t)\left[\frac{dr^{2}}{1 - kr^{2}} + r^{2}d\Omega^{2}\right],
\end{equation}
where $a(t)$ is the scale factor and the curvature constants $k$ are $-1, 0, +1$
for open, flat and close models of the universe respectively.\\\\
The Einstein's field equations (with $8\pi G = 1$ and $c = 1$) read as
\begin{equation}
\label{eq2}
R^{j}_{i} - \frac{1}{2}R\delta^{j}_{i} = - T^{j}_{i},
\end{equation}
where the symbols have their usual meaning and $T^{j}_{i}$ is the two fluid energy-momentum tensor consisting of
dark field and barotropic fluid.\\\\
In a co-moving coordinate system, Einstein's field equations (\ref{eq2}) for the line element (\ref{eq1}) lead to
\begin{equation}
\label{eq3} p_{tot} = -\left(2\frac{\ddot{a}}{a} + \frac{\dot{a}^{2}}{a^{2}} + \frac{k}{a^{2}}\right),
\end{equation}
and
\begin{equation}
\label{eq4}\rho_{tot} = 3\left(\frac{\dot{a}^{2}}{a^{2}} + \frac{k}{a^{2}}\right),
\end{equation}
where $p_{tot} = p_{m} + p_{D}$ and $\rho_{tot} = \rho_{m} + \rho_{D}$. Here $p_{m}$ and $\rho_{m}$ are pressure
and energy density of barotropic fluid and $p_{D}$ \& $\rho_{D}$ are pressure and energy density of dark fluid
respectively.\\\\
The Bianchi identity $G^{;j}_{ij} = 0$ leads to  $T^{;j}_{ij} = 0$ which yields
\begin{equation}
\label{eq5}\dot{\rho}_{tot} + 3\frac{\dot{a}}{a}\left(\rho_{tot} + p_{tot}\right)=0.
\end{equation}
The EoS of the barotropic fluid  and dark field are given by
\begin{equation}
\label{eq6}\omega_{m} = \frac{p_{m}}{\rho_{m}},
\end{equation}
and
\begin{equation}
\label{eq7}\omega_{D} = \frac{p_{D}}{\rho_{D}},
\end{equation}
respectively.\\

In the following sections we deal with two cases, (i) non-interacting two-fluid model and (ii) interacting two-fluid
model.
\section{Non-interacting two-fluid model}
First, we consider that two-fluids do not interact with each other. Therefore, the general form of
conservation equation (\ref{eq5}) leads us to writing the conservation equation for the dark and barotropic fluid
separately as,
\begin{equation}
\label{eq8}\dot{\rho}_{m} + 3\frac{\dot{a}}{a}\left(\rho_{m} + p_{m}\right) = 0,
\end{equation}
and
\begin{equation}
\label{eq9}\dot{\rho}_{D} + 3\frac{\dot{a}}{a}\left(\rho_{D} + p_{D}\right) = 0.
\end{equation}
Integration of Eq. (\ref{eq5}) leads to
\begin{equation}
\label{eq10}\rho_{m} = \rho_{0}a^{-3(1 + \omega_{m})},
\end{equation}
where $\rho_{0}$ is an integrating constant. By using Eq. (\ref{eq10}) in Eqs. (\ref{eq3}) and (\ref{eq4}), we
first obtain the $\rho_{D}$ and $p_{D}$ in term of scale factor $a(t)$
\begin{equation}
\label{eq11}\rho_{D} = 3\left(\frac{\dot{a}^{2}}{a^{2}} + \frac{k}{a^{2}}\right) - \rho_{0}a^{-3(1 + \omega_{m})}.
\end{equation}
\begin{figure}[htbp]
\centering
\includegraphics[width=8cm,height=8cm,angle=0]{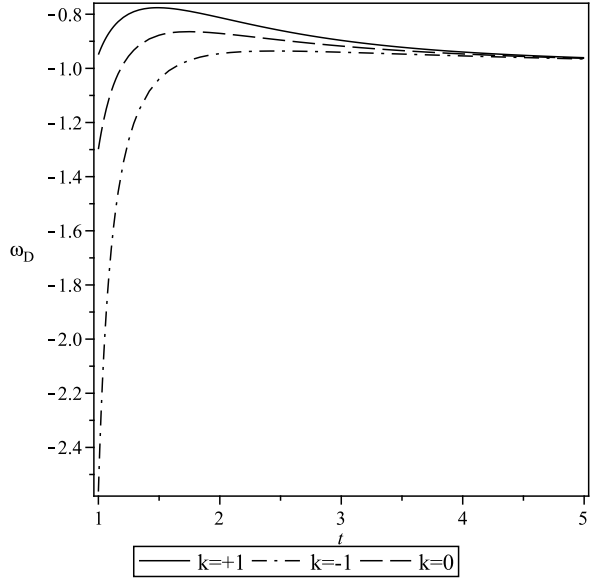}
\caption{The plot of EoS parameter vs $t$ for $\rho_{0}=10$ and $\omega_{m}=0.5$ }
\end{figure}
and
\begin{equation}
\label{eq12} p_{D} = -\left(2\frac{\ddot{a}}{a} + \frac{\dot{a}^{2}}{a^{2}} + \frac{k}{a^{2}}\right) -
\rho_{0}\omega_{m} a^{-3(1 + \omega_{m})}.
\end{equation}

Now we take following {\it ansatz} for the scale factor, where increase in term of time evolution
\begin{equation}
\label{eq13} a(t) = \sqrt{te^{t}}.
\end{equation}
The motivation to choose such scale factor is behind the fact that the universe is accelerated expansion at 
present and decelerated expansion in the past. Also, the transition redshift from deceleration expansion to 
accelerated expansion is about 0.5. Thus, in general, the DP is not a constant but time variable. By the above 
choice of scale factor yields a time dependent DP. \\\\ 
By using this scale factor in Eqs. (\ref{eq11}) and (\ref{eq12}), the $\rho_{D}$
and  $p_{D}$ are obtained as
\begin{equation}
\label{eq14} \rho_{D} = 3\left[\frac{(1+t)^{2}}{4t^{2}} + \frac{k}{te^{t}}\right] - 
\rho_{0}(te^{t})^{-\frac{3}{2}(1 + \omega_{m})},
\end{equation}
and
\begin{equation}
\label{eq15} p_{D} = -\left[\frac{3(1+t)^{2}-4}{4t^{2}} + \frac{k}{te^{t}} + 
\rho_{0}\omega_{m}(te^{t})^{-\frac{3}{2}(1 + \omega_{m})}\right],
\end{equation}
respectively. By using Eqs. (\ref{eq14}) and (\ref{eq15}) in Eq. (\ref{eq7}), we find the equation of state of
dark field in term of time as
\begin{equation}
\label{eq16}\omega_{D} = -\left[\frac{\frac{3(1+t)^{2}-4}{4t^{2}} + \frac{k}{te^{t}} + 
\rho_{0}\omega_{m}(te^{t})^{-\frac{3}{2}(1+\omega_{m})}}{3\left[\frac{(1+t)^{2}}{4t^{2}} + 
\frac{k}{te^{t}}\right] - \rho_{0}(te^{t})^{-\frac{3}{2}(1 + \omega_{m})}}\right].
\end{equation}
The behavior of EoS for DE in term of cosmic time $t$ is shown in Fig.1. It is observed that although for open,
close and flat universes the EoS parameter is an increasing function of time, the rapidity of its growth at the
early stage depends on the type the universe. Later on it tends to the same constant value independent of
the types of the universe. \\\\
The expressions for the matter-energy density $\Omega_{m}$ and dark-energy density $\Omega_{D}$ are given by
\begin{equation}
\label{eq17}\Omega_{m} = \frac{\rho_{m}}{3H^{2}} = \frac{4t^{2}}{3(1+t)^{2}}
\rho_{0}(te^{t})^{-\frac{3}{2}(1+\omega_{m})},
\end{equation}
and
\begin{equation}
\label{eq18}\Omega_{D} = \frac{\rho_{D}}{3H^{2}} = 1 + \frac{4kt}{3(1+t)^{2}e^{t}} - \frac{4t^{2}}{3(1+t)^{2}}
\rho_{0}(te^{t})^{-\frac{3}{2}(1 + \omega_{m})},
\end{equation}
respectively. Equations (\ref{eq17}) and (\ref{eq18}) reduce to
\begin{equation}
\label{eq19}\Omega = \Omega_{m} + \Omega_{D} = 1 + \frac{4kt}{3(1+t)^{2}e^{t}}.
\end{equation}
From the right hand side of Eq. (\ref{eq19}) it is clear that in flat universe ($k = 0$), $\Omega = 1$ and in open
universe ($k = -1$), $\Omega < 1$ and in close universe ($k = + 1$), $\Omega > 1$. But at late time we see for
all flat, open and close universes $\Omega \to 1$. This result is compatible with the observational results.
Since our model predicts a flat universe for large times and the present-day universe is very close to flat, the
derived model is also compatible with the observational results. The variation of density parameter with cosmic
time has been shown in Fig.2. \\\\
\begin{figure}[htbp]
\centering
\includegraphics[width=8cm,height=8cm,angle=0]{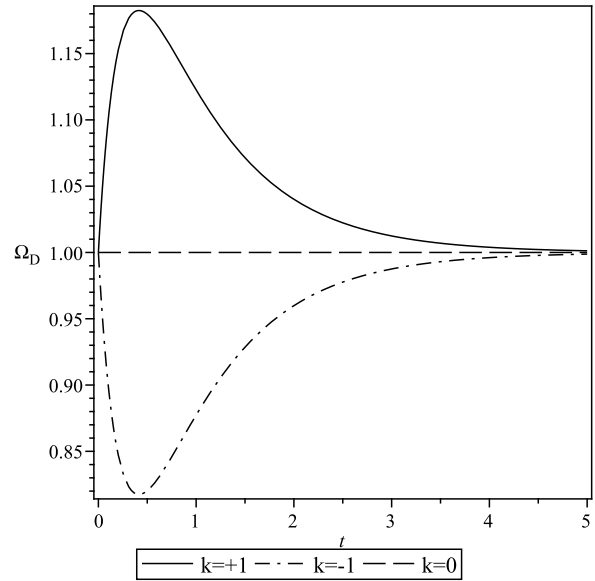}
\caption{The plot of density parameter ($\Omega$) vs $t$ for $n=3$ }
\end{figure}
We define the deceleration parameter $q$ as usual, i.e.
\begin{equation}
\label{eq20}q = - \frac{\ddot{a} a}{\dot{a}^{2}} = - \frac{\ddot{a}}{aH^{2}}.
\end{equation}
Using Eqs. (\ref{eq3}) and (\ref{eq4}), we may rewrite Eq. (\ref{eq20}) as
\begin{equation}
\label{eq21}q = \frac{1}{6H^{2}}\left[\rho_{m}(1 + 3\omega_{m}) + \rho_{D}(1 + 3\omega_{D})\right].
\end{equation}
On the other hand, using Eq. (\ref{eq13}) into Eq. (\ref{eq20}), we find
\begin{equation}
\label{eq22}q = \frac{2}{(1+t)^{2}}-1.
\end{equation}
From Eq. (\ref{eq22}), we observe that $q > 0 ~ ~ \mbox{for} ~ t < 0.41$ and $q < 0 ~ ~ \mbox{for} ~ t > 0.41$. 
This behavior of $q$ is clearly depicted in Fig.3 \\\\
\begin{figure}[htbp]
\centering
\includegraphics[width=8cm,height=8cm,angle=0]{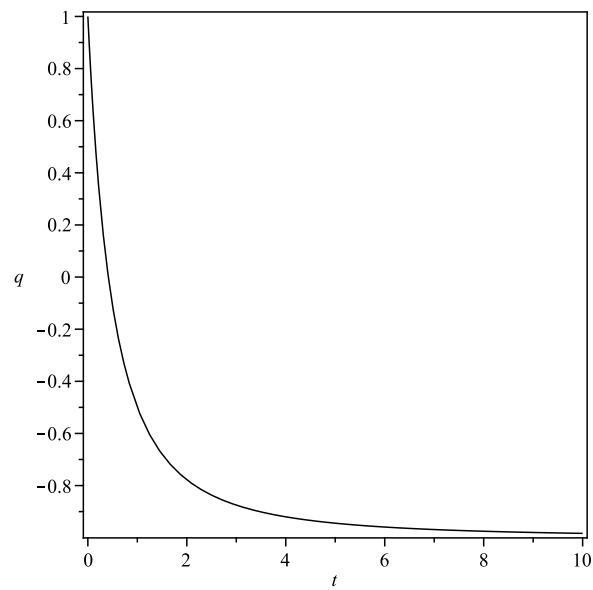}
\caption{The plot of deceleration parameter ($\Omega$) vs $t$ }
\end{figure}
A convenient method to describe models close to $\Lambda$ CDM is based on the cosmic jerk parameter $j$, a
dimensionless third derivative of the scale factor with respect to the cosmic time \cite{ref20}$-$\cite{ref23}. 
A deceleration-to-acceleration transition occurs for models with a positive value of $j_{0}$ and negative
$q_{0}$. Flat $\Lambda$ CDM models have a constant jerk $j = 1$. The jerk parameter in cosmology is defined
as the dimensionless third derivative of the scale factor with respect to cosmic time
\begin{equation}
\label{eq23} j(t) = \frac{1}{H^{3}}\frac{\dot{\ddot{a}}}{a}.
\end{equation}
and in terms of the scale factor to cosmic time
\begin{equation}
\label{eq24} j(t) = \frac{(a^{2}H^{2})^{''}}{2H^{2}},
\end{equation}
where the `dots' and `primes' denote derivatives with respect to cosmic time and scale factor, respectively.
The jerk parameter appears in the fourth term of a Taylor expansion of the scale factor around $a_{0}$,
\begin{equation}
\label{eq25} \frac{a(t)}{a_{0}} = 1 + H_{0}(t-t_{0}) - \frac{1}{2}q_{0}H_{0}^{2}(t-t_{0})^{2} +
\frac{1}{6}j_{0}H_{0}^{3}(t-t_{0})^{3} + O\left[(t-t_{0})^{4}\right],
\end{equation}
where the subscript $0$ shows the present value. One can rewrite Eq. (\ref{eq23}) as
\begin{equation}
\label{eq26} j(t) = q + 2q^{2} - \frac{\dot{q}}{H}.
\end{equation}
Equations (\ref{eq22}) and (\ref{eq26}) reduce to
\begin{equation}
\label{eq27} j(t) = \frac{t^{3}+3(t^{2}-t+1)}{(1+t)^{3}}.
\end{equation}
This value overlaps with the value $j\simeq2.16$ obtained from the combination of three kinematical data sets: the
gold sample of type Ia supernovae \cite{ref24}, the SNIa data from the SNLS project \cite{ref25}, and the x-ray
galaxy cluster distance measurements \cite{ref26} at $t\simeq 0.05$. 
\section{Interacting two fluids model}
Secondly, we consider the interaction between dark and barotropic fluids. For this purpose we
can write the continuity equations for dark fluid and barotropic fluids as
\begin{equation}
\label{eq28}\dot{\rho}_{m} + 3\frac{\dot{a}}{a}\left(\rho_{m} + p_{m}\right) = Q,
\end{equation}
and
\begin{equation}
\label{eq29}\dot{\rho}_{D} + 3\frac{\dot{a}}{a}\left(\rho_{D} + p_{D}\right) = -Q.
\end{equation}
The quantity $Q$ expresses the interaction between the dark components. Since we are interested in
 an energy transfer from the dark energy to dark matter, we consider $Q>0$. $Q>0$ ensures that the second
law of thermodynamics stands fulfilled \cite{ref27}. Here we emphasize that the continuity Eqs. (\ref{eq28}) and
(\ref{eq29}) imply that the interaction term ($Q$) should be  proportional to a quantity with units of inverse of
time. i.e $Q\propto \frac{1}{t}$. Therefore, a first and natural candidate can be the Hubble factor $H$ multiplied
with the energy density. Following Amendola et al. \cite{ref28} and Gou et al. \cite{ref29}, we consider
\begin{equation}
\label{eq30}Q = 3H \sigma \rho_{m},
\end{equation}
where $\sigma$ is a coupling constant. Using Eq. (\ref{eq30}) in Eq. (\ref{eq28}) and after integrating the
resulting equation, we obtain
\begin{equation}
\label{eq31}\rho_{m} = \rho_{0}a^{-3(1 + \omega_{m} - \sigma)}.
\end{equation}
By using Eq. (\ref{eq31}) in Eqs. (\ref{eq3}) and (\ref{eq4}), we again obtain the $\rho_{D}$ and $p_{D}$ in terms
of scale factor $a(t)$.
\begin{equation}
\label{eq32}\rho_{D} = 3\left(\frac{\dot{a}^{2}}{a^{2}} + \frac{k}{a^{2}}\right) -
\rho_{0}a^{-3(1 + \omega_{m} - \sigma)},
\end{equation}
and
\begin{equation}
\label{eq33} p_{D} = -\left(2\frac{\ddot{a}}{a} + \frac{\dot{a}^{2}}{a^{2}} + \frac{k}{a^{2}}\right) -
\rho_{0}(\omega_{m} - \sigma)a^{-3(1 + \omega_{m} - \sigma)},
\end{equation}
respectively. Putting the value of $a(t)$ from Eq. (\ref{eq13}) in Eqs. (\ref{eq32}) and (\ref{eq33}), we obtain
\begin{equation}
\label{eq34}\rho_{D} = 3\left[\frac{(1+t)^{2}}{4t^{2}}+\frac{k}{te^{t}}\right]-\rho_{0}(te^{t})^
{-\frac{3}{2}(1 + \omega_{m} - \sigma)},
\end{equation}
and
\begin{equation}
\label{eq35}p_{D} = -\left[\frac{3(1+t)^{2}-4}{4t^{2}}+\frac{k}{te^{t}}+\rho_{0}\omega_{m}(te^{t})^
{-\frac{3}{2}(1 + \omega_{m} - \sigma)}\right],
\end{equation}
respectively. Using Eqs. (\ref{eq34}) and (\ref{eq35}) in Eq. (\ref{eq7}), we can find the EoS parameter of dark
field as
\begin{equation}
\label{eq36}\omega_{D} = -\left[\frac{\frac{3(1+t)^{2}-4}{4t^{2}}+\frac{k}{te^{t}}+\rho_{0}\omega_{m}(te^{t})^
{-\frac{3}{2}(1 + \omega_{m} - \sigma)}}{3\left[\frac{(1+t)^{2}}{4t^{2}}+\frac{k}{te^{t}}\right]-\rho_{0}(te^{t})^
{-\frac{3}{2}(1 + \omega_{m} - \sigma)}}\right].
\end{equation}
The behavior of EoS in term of cosmic time $t$ is shown in Fig.4. It is observed that like the minimal
coupling case, the EoS parameter is an increasing function of time for all close, open and flat universes, 
the rapidity of its increase at the early stage depends on the type of universe. At the later stage of evolution it 
tends to the same constant value independent of the types of the Universe. The EoS parameter of DE begins in 
phantom region and tends to $-1$ (cosmological constant. \\\\
\begin{figure}[htbp]
\centering
\includegraphics[width=8cm,height=8cm,angle=0]{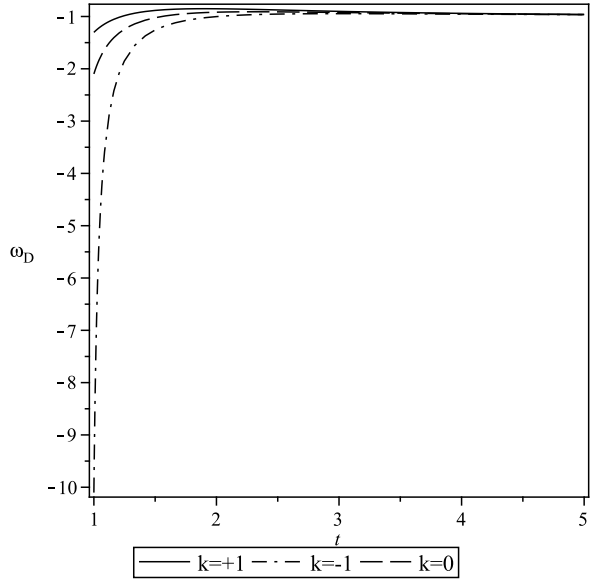}
\caption{The plot of EoS parameter vs $t$ for $\rho_{0} = 10, \omega_{m} = 0.5$ and $\sigma=0.3$ }
\end{figure}
The expressions for the matter-energy density $\Omega_{m}$ and dark-energy density $\Omega_{D}$ are given by
\begin{equation}
\label{eq37}\Omega_{m} = \frac{\rho_{m}}{3H^{2}} = \frac{4t^{2}}{3(1+t)^{2}}
\rho_{0}(te^{t})^{-3n(1+\omega_{m}-\sigma)},
\end{equation}
and
\begin{equation}
\label{eq38}\Omega_{D} = \frac{\rho_{D}}{3H^{2}} = 1 + \frac{4kt}{3(1+t)^{2}e^{t}}-\frac{4t^{2}}{3(1+t)^{2}}
\rho_{0}(te^{t})^{-3n(1 + \omega_{m} - \sigma)},
\end{equation}
respectively. From Eqs. (\ref{eq37}) and (\ref{eq38}), we obtain
\begin{equation}
\label{eq39}\Omega = \Omega_{m} + \Omega_{D} = 1 + \frac{4kt}{3(1+t)^{2}e^{t}},
\end{equation}
which is the same as Eq. (\ref{eq19}). Therefore, we observe that in the interacting case the density parameter 
has the same properties as in the non-interacting case. The expressions for deceleration parameter and jerk 
parameter are also the same as in the case of non-interacting case.\\\\
Studying the interaction between the dark energy and ordinary matter will open a possibility of detecting the dark
energy. It should be pointed out that evidence was recently provided by the Abell Cluster A586 in support of the
interaction between dark energy and dark matter \cite{ref30, ref31}. We observe that in the non-interacting case only
open and flat universes can cross the phantom region whereas in interacting case all open, flat and close 
universes can cross phantom region. 
\section{Concluding Remarks}
In summary, we have studied the system of two-fluid within the scope of a spatially flat and isotropic FRW model. 
The role of two-fluid minimally or directly coupled in the evolution of the dark energy parameter has been
investigated. In doing so the scale factor is taken to be an exponential law function of time. It is concluded 
that in the non-interacting case only open and flat universes cross the phantom region whereas in the interacting 
case all three universes can cross the phantom region. 
\section*{Acknowledgments}
One of the authors (A. Pradhan) would like to thank the Laboratory of Information Technologies, Joint Institute for
Nuclear Research, Dubna, Russia for providing facility and support, where a part of this work was carried out. The 
authors thank the anonymous referees for valuable comments.

\end{document}